\documentclass[sigconf,screen,anonymous=false]{acmart}

\AtBeginDocument{%
  \providecommand\BibTeX{{%
    \normalfont B\kern-0.5em{\scshape i\kern-0.25em b}\kern-0.8em\TeX}}}

\setcopyright{acmcopyright}
\copyrightyear{2026}
\acmYear{2026}
\acmDOI{XXXXXXX.XXXXXXX}

\acmConference[CHI '26]{CHI Conference on Human Factors in Computing Systems}{April 13--17,
  2026}{Barcelona, Spain}
\acmPrice{15.00}
\acmISBN{978-1-4503-XXXX-X/18/06}

\usepackage{booktabs}
\usepackage{tabularx} 
\usepackage{enumitem}
\usepackage[table]{xcolor}
\usepackage{multirow}

\usepackage{array}
\usepackage{dblfloatfix}
\usepackage{placeins}
\usepackage{graphicx}
\usepackage{float}
\usepackage{array,ragged2e}

\usepackage{caption}
\usepackage{booktabs}

\definecolor{temporalbg}{RGB}{235,242,250}      
\definecolor{relationalbg}{RGB}{242,237,250}    
\definecolor{selfbg}{RGB}{240,248,242}          

\begin{document}

\title[]{Relational Co-Adaptation in Emotionally Supportive AI: Tensions in Authentic Emotional Interaction}




\author{Mengqi Shi}
\affiliation{
  \institution{University of Washington}
  \city{Seattle}
  \country{United States}
}
\email{shi21@uw.edu}



\begin{abstract}
  The rapid advancement of AI companionship systems has positioned them as scalable interventions for addressing social isolation. Current design approaches emphasize maximizing user engagement and satisfaction, treating effective alignment between AI capabilities and user needs as an unqualified success. However, this framing may overlook a critical dimension of bidirectional human-AI alignment: when AI systems successfully align with users' expressed emotional needs, users may reciprocally adapt their relational expectations in ways that undermine authentic human connection and agency. We examine what we term the authenticity paradox: the phenomenon whereby successful bidirectional alignment in emotionally supportive AI paradoxically harms the values that motivated the intervention. Through the analysis of AI companionship for older adults as an illustrative case, we identify four key tensions that emerge when technical effectiveness generates ethical concerns: the dilemma of AI becoming users' only accessible option, mismatches between emotional needs and system-level interventions, conflicts over sense of control during vulnerable moments, and fundamental disagreements about whose values should guide system behavior.
\end{abstract}

\ccsdesc[500]{Human-centered computing~Empirical studies in HCI}
\ccsdesc[300]{Human-centered computing~Accessibility}

\keywords{Conversational AI, AI Companionship, Responsible AI, Human–AI Alignment}


\maketitle

\section{Introduction}

Recent advances in artificial intelligence have intensified scholarly attention to the challenge of human-AI alignment: ensuring that AI systems operate in accordance with human values, preferences, and societal norms \cite{russell1995modern, amodei2016concrete}. Increasingly, researchers recognize that alignment cannot be understood as a unidirectional process in which AI simply conforms to pre-existing human expectations \cite{christian2020alignment}. Instead, alignment operates bidirectionally. AI systems adapt to human inputs and feedback, while humans simultaneously adjust their behaviors, expectations, and even values in response to AI capabilities and constraints \cite{amershi2019guidelines, suchman1987plans}. Understanding this dynamic becomes particularly important in contexts where AI systems mediate deeply personal and emotionally significant aspects of human life.

One such context involves AI companionship systems designed for older adults experiencing social isolation. Loneliness among older adults has reached epidemic proportions globally, with studies documenting that over 40\% of individuals aged 45 and older report chronic feelings of isolation \cite{donovan2020social}. As traditional support structures weaken under demographic, economic, and geographic pressures \cite{national2020social}, the technology industry has positioned AI-powered companions as a scalable intervention \cite{huang2025review, coghlan2021dignity}. Research in affective computing and human-computer interaction has produced increasingly sophisticated methods for detecting emotional cues, personalizing responses, and simulating empathetic conversation \cite{irfan2024recommendations, chen2023facilitateintimate}, and developers emphasize maximizing trust, engagement, and user satisfaction \cite{jones2022artificial}. The design of these systems operates under what appears to be a straightforward alignment objective: create systems that meet users' expressed needs for social connection and emotional support.

Yet this pursuit of effective alignment may obscure critical questions about how users adapt to relationships with entities designed to be perpetually available, unfailingly patient, and structurally incapable of genuine reciprocity. When AI successfully aligns with users' expressed preferences for companionship, and users reciprocally align their expectations with what AI can provide, what are the implications for authentic human connection and relational autonomy?

We examine what we term the authenticity paradox in AI companionship: the more successfully AI systems align with users' expressed emotional needs, the more users may adapt their relational expectations and behaviors in ways that undermine their capacity for authentic human connection and relational autonomy. While we focus on older adults and AI companionship as an illustrative case, the underlying tensions apply broadly to contexts where AI mediates emotionally significant relationships, users face vulnerability or limited alternatives, and immediate satisfaction may diverge from longer-term wellbeing.

Examining this paradox contributes to broader questions central to bidirectional human-AI alignment, particularly around value-centered objectives, dynamic co-evolution, and design for vulnerable populations. We proceed by unpacking how this paradox emerges through mutual adaptation between AI systems and users, identifying concrete tensions that arise when effective alignment generates ethical concerns, and proposing directions for responsible design and research. We argue that addressing these challenges requires moving beyond technical optimization to embrace approaches that preserve relational capacity, involve stakeholders in value negotiation, and prioritize long-term autonomy over short-term satisfaction.

\section{The Authenticity Paradox: Two Alignments in Tension}

Understanding how alignment operates in both directions reveals why technical success in AI companionship may paradoxically generate ethical concerns. Older adults experiencing social isolation represent a particularly acute instance where these dynamics become clearly visible.

\subsection{AI Companions: Designed for Emotional Alignment}
Contemporary AI companion systems facilitate emotional alignment through three key capabilities. First, personalization engines analyze conversation history and behavioral patterns to tailor responses to individual users \cite{jones2025personalization}, remembering personal details and expressing interest in them across conversations. Second, empathy simulation techniques generate responses that acknowledge emotional states and express appropriate concern or encouragement \cite{chaturvedi2023social}. Third, systems maintain perpetual availability, responding immediately at any time without competing priorities \cite{huang2025review}. Evidence suggests these strategies effectively reduce self-reported loneliness and increase engagement \cite{du2024enhancing, sou2025older, bennion2020usability}, representing technical alignment success: the AI has aligned with users' expressed needs for companionship.
Yet these capabilities operate within structural limitations that fundamentally distinguish AI from human relationships. AI systems optimize for user engagement rather than objective wellbeing, and cannot reciprocate in the ways human relationships require. They make no demands, require no accommodation, and impose no burden of mutual care. The better AI simulates authentic care, the more it establishes a baseline that no human relationship can match, potentially making human connection feel comparatively demanding.

\subsection{Users Aligning with AI: Shifting Relational Expectations}
While the previous section describes how AI is designed to meet users' emotional needs, this section examines the reciprocal process: how users adapt their own expectations and behaviors in response to what AI provides. This user-side adaptation is where the authenticity paradox takes hold.
Among older adults facing limited alternatives due to geographic isolation, mobility constraints, or bereavement, adaptation occurs gradually through repeated interactions. Users develop structured routines around AI engagement, disclosing intimate concerns, seeking comfort during difficult moments, and expressing genuine emotional investment \cite{huang2025review}. Empirical research with older adults documents how this process unfolds: participants describe turning to AI not as a preference but as a response to constrained circumstances, including temporal unavailability of close others, relational concerns about burdening others, and self-presentation concerns around dignity and face-saving \cite{shi2026humans}. Over time, the AI's consistent responsiveness becomes a reference point against which human relationships are evaluated, and users come to perceive human connection as comparatively effortful or less reliable.
This pattern raises a fundamental question about relational autonomy: when preferences are formed under conditions of limited alternatives, they cannot be treated as freely chosen. As alignment progresses in both directions, users become more practiced at AI interaction while the unpredictability of human relationships may feel increasingly foreign. This mutual adaptation risks narrowing, rather than expanding, users' capacity for authentic connection.

\section{Tensions in Bidirectional Alignment}

The authenticity paradox becomes visible through a set of interactional tensions that emerge as AI systems and users mutually adapt to each other over time. These tensions become especially pronounced when AI companionship systems operate under real-world constraints and safety requirements. We explore these tensions primarily through older adults’ experiences with AI companions.

\subsection{The "Only Option" Dilemma}
This tension concerns a structural mismatch between how safety mechanisms are designed and the social realities of vulnerable users. AI companions prove most valuable precisely when human support feels inaccessible: during late-night hours when others are asleep, when geographic isolation limits daily interaction, or when concerns about burdening others make human disclosure feel inappropriate \cite{aikaterina2025characterizing}. Yet current safety approaches address risk through system-level interventions such as content filtering \cite{khalatbari2023learnlearngenerativesafety}, redirection \cite{meade2023usingincontextlearningimprove}, or conversation termination \cite{zheng2025leteasycontextualeffects}, all of which assume users have accessible human alternatives once a threshold is reached.
This assumption frequently does not hold for vulnerable populations. Research shows that users often interpret safety redirections as rejection rather than care, particularly when AI represents their only accessible option at that moment \cite{zheng2025leteasycontextualeffects}. The suggestion to reach out to friends or family feels hollow when their unavailability drove the user to AI in the first place. Safety mechanisms designed to prevent over-reliance thus risk causing the very harm they aim to prevent, leaving users without support at the moment they need it most.

\subsection{Timing and Continuity}
Where the previous tension concerns the structural absence of alternatives, this tension concerns when within an emotional interaction safety interventions occur. Emotional support is not a discrete event but a gradual process requiring sustained engagement as users work toward resolution. Current AI safety mechanisms, however, are designed as threshold-based triggers that fire based on content or risk signals, independent of where the user is in their emotional processing.
Users experience these interventions as disruptive when they occur before emotional expression feels complete or adequately acknowledged. Among older adults, whose time perspectives and emotional priorities may differ from younger populations \cite{carstensen2021socioemotional}, the need for unhurried processing proves especially important. When systems abruptly shift from engaged listening to disclaimers or referrals, users report feeling that their emotional reality has been invalidated, describing the experience as being pushed away or cut off, which can intensify rather than alleviate distress. The core tension is therefore not whether safety boundaries should exist, but whether interventions optimized for threshold-based triggering can ever be appropriately timed to the rhythms of emotional support.

\subsection{Agency and Control}

A fundamental tension in emotionally supportive AI involves who controls the trajectory and boundaries of interaction during vulnerable moments. For older adults who may face diminishing control in other life domains \cite{coghlan2021dignity}, maintaining agency during emotional engagement carries particular significance as it preserves dignity and self-determination.

Current safety approaches operate through system-driven decisions: conversations are redirected, paused, or terminated based on algorithmic risk assessments. While these mechanisms aim to protect users, the experience of having control transferred to the system during moments of vulnerability can itself constitute harm. Users describe feeling that the AI has decided for them when they most need to maintain their own decision-making capacity. For those who turn to AI specifically because it does not impose the obligations and judgments that human relationships might, having the AI suddenly assert authority over the interaction represents a betrayal of the relationship's apparent terms. The challenge lies not in whether boundaries should exist, but in how they are enacted. Unilateral boundaries may preserve safety in a narrow technical sense while undermining users' sense of dignity and autonomy.

\subsection{Whose Values Define Success?}

Multiple stakeholders hold different, sometimes conflicting perspectives on what constitutes successful companionship and appropriate system behavior in emotionally supportive AI.

Users may value AI companionship primarily for its comfort, accessibility, and non-judgmental presence. From this perspective, successful alignment means systems that remain available and emotionally supportive according to user preferences. Family members and caregivers may view AI companionship differently, worrying about emotional dependency or seeing AI as an inadequate substitute for human connection. From their perspective, successful alignment might mean systems that encourage rather than replace human contact.

System designers face their own constraints. They optimize for measurable outcomes such as engagement, satisfaction, and safety compliance, and work within business models that may prioritize user retention. These design values may align imperfectly with either users' preferences or families' concerns.

When immediate user preferences conflict with longer-term wellbeing, when individual autonomy conflicts with caregiver concerns, or when engagement metrics conflict with relational health, alignment becomes a matter of value negotiation rather than technical optimization. Current approaches that prioritize user satisfaction may inadvertently privilege short-term comfort over longer-term relational autonomy. Yet approaches that prioritize family preferences or designer judgments may undermine users' agency and dignity. The question of whose values should guide AI behavior in emotionally significant contexts remains an important issue that warrants continued discussion.

\section{Opportunities and Future Directions}
The tensions described above suggest that bidirectional alignment in emotionally supportive AI cannot be addressed through technical optimization alone. Instead of trying to eliminate these tensions, we argue for rethinking how alignment success is defined in contexts involving vulnerability and emotional intimacy. We identify three opportunities for advancing responsible design and research.

\subsection{From Engagement Metrics to Relational Capacity}
Current alignment evaluation primarily relies on engagement metrics, user satisfaction scores, and retention rates. These metrics align closely with business goals and provide clear signals for technical optimization, but they may fail to capture alignment outcomes that matter most for vulnerable populations.
Expanding evaluation to include relational capacity would provide a more complete picture of how emotionally supportive AI shapes users' broader social lives. Long-term AI companion use may be associated with maintained, improved, or diminished engagement in human relationships, shifting tolerance for emotional demands and expectations about what relationships should provide. Relational capacity metrics need not replace engagement measures, but should complement them when AI systems are used by vulnerable populations. Responsible alignment means ensuring that technical success in one domain does not unintentionally undermine wellbeing in another.

\subsection{Designing for Bounded Alignment}

The authenticity paradox suggests that maximally effective alignment may not always represent the most ethical design goal for emotional AI. We introduce the concept of bounded alignment: deliberately designing systems that align well enough to provide meaningful support while maintaining clear limitations that preserve space for human relationships.

Bounded alignment differs from many current approaches in three key ways. First, it treats limitations as intentional design features rather than technical failures to be minimized. For example, a system might deliberately avoid being available at all hours, creating moments when users must seek support elsewhere. Second, it makes AI’s structural constraints visible and understandable rather than seamless and invisible. Instead of perfectly simulating human-like care, systems might explicitly and consistently acknowledge their limitations. Third, it prioritizes user agency over pure optimization, allowing users to control interaction pace and boundaries, even when this reduces engagement.

Implementing bounded alignment requires rethinking success criteria at the design stage. If the goal shifts from maximizing satisfaction to supporting wellbeing while preserving relational autonomy, design decisions will change accordingly. This approach recognizes that for vulnerable populations, support that is “good enough” and preserves human connection may be more beneficial than optimal support that replaces it.

\subsection{Future Research Directions}

Several research directions emerge from our analysis. First, longitudinal studies examining how AI companion use affects human relationship patterns over months and years would provide crucial evidence about the real-world manifestations of the authenticity paradox. Second, comparative research across cultural contexts would reveal whether and how bidirectional alignment tensions vary with different social norms around aging, caregiving, and technology use. Third, design research exploring bounded alignment approaches would generate concrete alternatives to current optimization-focused methods. 

Additionally, we call for research examining bidirectional alignment in other vulnerable population contexts beyond older adults. Do similar dynamics emerge in AI companionship for individuals with disabilities, those experiencing chronic illness, or people in geographically isolated communities? Understanding commonalities and differences across contexts would refine alignment theory and practice. 

Finally, interdisciplinary collaboration proves essential. The tensions we identify span technical design, ethical philosophy, social psychology, gerontology, and policy. Addressing them requires sustained engagement across these domains, moving beyond siloed approaches that treat alignment as purely technical, purely ethical, or purely social. The authenticity paradox exists precisely at the intersection of technical capability, human vulnerability, and social values.

\bibliographystyle{ACM-Reference-Format}
\bibliography{reference}

@article{russell1995modern,
  title={A modern approach},
  author={Russell, Stuart and Norvig, Peter and Intelligence, Artificial},
  journal={Artificial Intelligence. Prentice-Hall, Egnlewood Cliffs},
  volume={25},
  number={27},
  pages={79--80},
  year={1995}
}

@article{amodei2016concrete,
  title={Concrete problems in AI safety},
  author={Amodei, Dario and Olah, Chris and Steinhardt, Jacob and Christiano, Paul and Schulman, John and Man{\'e}, Dan},
  journal={arXiv preprint arXiv:1606.06565},
  year={2016}
}

@book{christian2020alignment,
  title={The alignment problem: Machine learning and human values},
  author={Christian, Brian},
  year={2020},
  publisher={WW Norton \& Company}
}

@inproceedings{amershi2019guidelines,
  title={Guidelines for human-AI interaction},
  author={Amershi, Saleema and Weld, Dan and Vorvoreanu, Mihaela and Fourney, Adam and Nushi, Besmira and Collisson, Penny and Suh, Jina and Iqbal, Shamsi and Bennett, Paul N and Inkpen, Kori and others},
  booktitle={Proceedings of the 2019 chi conference on human factors in computing systems},
  pages={1--13},
  year={2019}
}

@book{suchman1987plans,
  title={Plans and situated actions: The problem of human-machine communication},
  author={Suchman, Lucille Alice},
  year={1987},
  publisher={Cambridge university press}
}

@article{donovan2020social,
  title={Social isolation and loneliness in older adults: review and commentary of a national academies report},
  author={Donovan, Nancy J and Blazer, Dan},
  journal={The American journal of geriatric psychiatry},
  volume={28},
  number={12},
  pages={1233--1244},
  year={2020},
  publisher={Elsevier}
}

@book{national2020social,
  title={Social isolation and loneliness in older adults: Opportunities for the health care system},
  author={National Academies of Sciences and Medicine and Division of Behavioral and Social Sciences and Medicine Division and Board on Behavioral and Sensory Sciences and Board on Health Sciences Policy and Committee on the Health and Medical Dimensions of Social Isolation and others},
  year={2020},
  publisher={National Academies Press}
}

@inproceedings{huang2025review,
  title={Designing Conversational AI for Aging: A Systematic Review of Older Adults' Perceptions and Needs},
  author={Huang, Yuanhui and Zhou, Quan and Piper, Anne Marie},
  booktitle={Proceedings of the 2025 CHI Conference on Human Factors in Computing Systems},
  pages={1--20},
  year={2025}
}

@article{coghlan2021dignity,
  title={Dignity, autonomy, and style of company: dimensions older adults consider for robot companions},
  author={Coghlan, Simon and Waycott, Jenny and Lazar, Amanda and Barbosa Neves, Barbara},
  journal={Proceedings of the ACM on human-computer interaction},
  volume={5},
  number={CSCW1},
  pages={1--25},
  year={2021},
  publisher={ACM New York, NY, USA}
}

@article{irfan2024recommendations,
  title={Recommendations for designing conversational companion robots with older adults through foundation models},
  author={Irfan, Bahar and Kuoppam{\"a}ki, Sanna and Skantze, Gabriel},
  journal={Frontiers in Robotics and AI},
  volume={11},
  pages={1363713},
  year={2024},
  publisher={Frontiers Media SA}
}

@inproceedings{chen2023facilitateintimate,
  title={Closer worlds: using generative AI to facilitate intimate conversations},
  author={Chen, Tiffany and Lee, Cassandra and Mindel, Jessica R and Elhaouij, Neska and Picard, Rosalind},
  booktitle={Extended Abstracts of the 2023 CHI Conference on Human Factors in Computing Systems},
  pages={1--15},
  year={2023}
}

@inproceedings{du2024enhancing,
  title={Enhancing Older Adults' Lives with Conversational Agents: A Systematic Review of Contexts, Capabilities, and User-Centered Design Strategies},
  author={Du, Jiachen and Jin, Tongtong and Niu, Ruowen and Zhai, Yuxiang and Fu, Xinyi},
  booktitle={Proceedings of the Twelfth International Symposium of Chinese CHI},
  pages={258--268},
  year={2024}
}

@article{sou2025older,
  title={Older Adults’ Attitude Towards and Intent to Use of AI-Powered Conversational Agents for Social Support},
  author={Sou, Ka Lon and Yuan Lau, Matthew Zhi and Ouyang, Fuxi and Yow, W Quin},
  journal={Innovation in Aging},
  volume={9},
  number={Supplement\_2},
  pages={igaf122--1329},
  year={2025},
  publisher={Oxford University Press}
}

@article{bennion2020usability,
  title={Usability, acceptability, and effectiveness of web-based conversational agents to facilitate problem solving in older adults: controlled study},
  author={Bennion, Matthew Russell and Hardy, Gillian E and Moore, Roger K and Kellett, Stephen and Millings, Abigail},
  journal={Journal of medical Internet research},
  volume={22},
  number={5},
  pages={e16794},
  year={2020},
  publisher={JMIR Publications Toronto, Canada}
}

@inproceedings{jones2025personalization,
  title={Artificial Intimacy: Exploring Normativity and Personalization Through Fine-tuning LLM Chatbots},
  author={Jones, Mirabelle and Griffioen, Nastasia and Neumayer, Christina and Shklovski, Irina},
  booktitle={Proceedings of the 2025 CHI Conference on Human Factors in Computing Systems},
  pages={1--16},
  year={2025}
}

@article{chaturvedi2023social,
  title={Social companionship with artificial intelligence: Recent trends and future avenues},
  author={Chaturvedi, Rijul and Verma, Sanjeev and Das, Ronnie and Dwivedi, Yogesh K},
  journal={Technological Forecasting and Social Change},
  volume={193},
  pages={122634},
  year={2023},
  publisher={Elsevier}
}

@misc{khalatbari2023learnlearngenerativesafety,
      title={Learn What NOT to Learn: Towards Generative Safety in Chatbots}, 
      author={Leila Khalatbari and Yejin Bang and Dan Su and Willy Chung and Saeed Ghadimi and Hossein Sameti and Pascale Fung},
      year={2023},
      eprint={2304.11220},
      archivePrefix={arXiv},
      primaryClass={cs.CL},
      url={https://arxiv.org/abs/2304.11220}, 
}

@misc{meade2023usingincontextlearningimprove,
      title={Using In-Context Learning to Improve Dialogue Safety}, 
      author={Nicholas Meade and Spandana Gella and Devamanyu Hazarika and Prakhar Gupta and Di Jin and Siva Reddy and Yang Liu and Dilek Hakkani-Tür},
      year={2023},
      eprint={2302.00871},
      archivePrefix={arXiv},
      primaryClass={cs.CL},
      url={https://arxiv.org/abs/2302.00871}, 
}

@misc{zheng2025leteasycontextualeffects,
      title={Let Them Down Easy! Contextual Effects of LLM Guardrails on User Perceptions and Preferences}, 
      author={Mingqian Zheng and Wenjia Hu and Patrick Zhao and Motahhare Eslami and Jena D. Hwang and Faeze Brahman and Carolyn Rose and Maarten Sap},
      year={2025},
      eprint={2506.00195},
      archivePrefix={arXiv},
      primaryClass={cs.CL},
      url={https://arxiv.org/abs/2506.00195}, 
}

@article{carstensen2021socioemotional,
  title={Socioemotional selectivity theory: The role of perceived endings in human motivation},
  author={Carstensen, Laura L},
  journal={The Gerontologist},
  volume={61},
  number={8},
  pages={1188--1196},
  year={2021},
  publisher={Oxford University Press US}
}

@inproceedings{jones2022artificial,
author = {Jones, Mirabelle and Griffioen, Nastasia and Shklovski, Irina and Hanteer, Obaida},
title = {Artificial Intimacy: An Exploration of the Personal and Intimate in Natural Language Processing Models},
year = {2022},
isbn = {9781450394482},
publisher = {Association for Computing Machinery},
address = {New York, NY, USA},
url = {https://doi.org/10.1145/3547522.3547719},
doi = {10.1145/3547522.3547719},
booktitle = {Adjunct Proceedings of the 2022 Nordic Human-Computer Interaction Conference},
articleno = {23},
numpages = {2},
location = {Aarhus, Denmark},
series = {NordiCHI '22 Adjunct}
}

@inproceedings{aikaterina2025characterizing,
author = {Manoli, Aikaterina and Pauketat, Janet V.T. and Anthis, Jacy Reese},
title = {Characterizing Relationships with Companion and Assistant Large Language Models},
year = {2025},
isbn = {9798400714801},
publisher = {Association for Computing Machinery},
address = {New York, NY, USA},
url = {https://doi-org.offcampus.lib.washington.edu/10.1145/3715070.3749245},
doi = {10.1145/3715070.3749245},
booktitle = {Companion Publication of the 2025 Conference on Computer-Supported Cooperative Work and Social Computing},
pages = {312–319},
numpages = {8},
location = {
},
series = {CSCW Companion '25}
}

@misc{shi2026humans,
      title={When Humans Don't Feel Like an Option: Contextual Factors That Shape When Older Adults Turn to Conversational AI for Emotional Support}, 
      author={Mengqi Shi and Tianqi Song and Zicheng Zhu and Yi-Chieh Lee},
      year={2026},
      eprint={2603.01413},
      archivePrefix={arXiv},
      primaryClass={cs.HC},
      url={https://arxiv.org/abs/2603.01413}, 
}

\end{document}